\documentstyle[12pt]  {article}
\begin{document}
\title{\bf Thermo--hydrodynamics \\ As a Field Theory}
\author{Jacek Jezierski\\
Department of Mathematical Methods in
Physics, \\ University of Warsaw,
ul. Ho\.za 74, 00-682 Warsaw,
Poland \and Jerzy Kijowski\\
Institute for Theoretical Physics, Polish Academy of Sciences,
\\ Aleja Lotnik\'ow 32/46, 02-668 Warsaw, Poland}

\date{}

\maketitle
 
\newcommand{\be}{\begin{equation}}
\newcommand{\ee}{\end{equation}} 
\newcommand{\ten}[3]{{#1}^{#2}_{\ #3}}
\newcommand{\wek}[2]{{#1}^{#2}}
\newcommand{\kowek}[2]{{#1}_{#2}}
\newcommand{\base}[2]{{{\partial}\over {\partial
{#1}^{#2}}}} 
\newcommand{\diff}[2]{\frac{\partial{#1}}{\partial{#2}}} 

\begin{abstract}
The field theoretical description of thermo--hydrodynamics is
given. It is based on the duality between the physical
space--time and the ``material space--time'' which we construct
here. The material space appearing in a natural way in the
canonical formulation of the hydrodynamics is completed with a
material time playing role of the field potential for temperature.
Both Lagrangian and Hamiltonian formulations, the canonical
structure, Poisson bracket, N\"other theorem and conservation laws
are discussed.
\end{abstract}
 
\section  {Introduction}
The simplest way to recognize the canonical structure of the
 electrodynamics consists  in introducing the electrodynamical
 potentials. This way the homogeneous Maxwell equations become
 automatically satisfied and non--homogeneous ones can be 
 interpreted as 2-nd order differential equations for the 
 potentials. The equations can be derived from the first order,
 non--degenerate variational principle. This makes the life easy: the
 conservations laws can be deduced from the N\"other theorem and the
 canonical structure is immediately given with ``Dirac delta''
 Poisson brackets between the potentials and their canonical momenta.
 The entire system becomes in a natural way an infinite dimensional
 hamiltonian system. Although it is possible to reduce the above
 structure rewriting it in terms of the electrodynamic fields only,
 without the use of potentials, but the resulting structure of the
 ``non--canonical'' Poisson brackets is very complicated and --
 for some applications -- useless.

 The situation in hydrodynamics is very much similar. Introducing
 the material space and describing consequently the configuration of 
the fluid as a diffeomorphism between the material space and the 
physical space makes the theory formally similar to the 
electrodynamics. The matter continuity equation is automatically
satisfied and the Euler equations can be interpreted as
the second order differential equations for the field potentials.
The equations can be derived from the first order, 
non--degenerate variational principle (see\cite{KPT}). The
corresponding hamiltonian theory is based on the canonical
(``Dirac delta'') Poisson structure.

It is very interesting that also the thermodynamics can be
formulated this way. We introduce the potential for the temperature 
which can be interpreted as a ``material time'' (see\cite{Kij-Tulcz2})
and define the temperature as a ratio between the  two different times 
(the material one and the physical one). This way the thermo--hydrodynamics
becomes a lagrangian field theory which fits also very well into
the framework of the infinite dimensional hamiltonian systems.

Having at our disposal the two time variables, we are free
to choose one of them as a independent variable of the theory
which parameterizes the physical events, and the other one as a
field potential for the temperature. 
These two choices correspond to the ``energy picture'' and the
``entropy picture'' in thermodynamics. The transformation between
the two pictures exchanges the role of the momenta canonically
conjugate to the two ``time variables'', i.e. between the energy
and the entropy.

The above purely phenomenological approach has also a nice microscopic
interpretation which we give at the end of the paper.
Some applications to the non--conservative phenomena can be found in 
\cite{ComptesRendues}.

\section {Barotropic fluid: Lagrange picture}

We assume that the collection of all the points of the
material can be organized in a smooth, 3--dimensional 
differential manifold $Y$ which we call the {\it material
space}. The points of $Y$ label the particles of the material. We
introduce coordinates $y^a (a=1,2,3)$  in the material space. Together with the
physical time $t$ parameterizing the time axis $U$ they form the
independent variables of our theory. It is useful to denote
$y^0:=t$ and to consider coordinates $(y^{\alpha})$, with
$\alpha=0,1,2,3$, in the space  $Y \times U$  of independent
variables of the theory.

To describe different types of materials we have to equip the
material space $Y$ with different geometric structures.
For example, the elastodynamics
needs the metric (riemannian) structure in $Y$. For
purposes of hydrodynamics, however, the only structure we need is
the volume structure. This means that there is in $Y$ a
differential 3--form (a scalar density) $r$ which measures the
quantity of the fluid (e.g. in moles) when integrated over
regular domains of $Y$. In terms of coordinates $y^a$ the
volume form $r$ can be written as follows: 
\be
r=h(y)dy^1 \wedge dy^2 \wedge dy^3 \ ,  \label{r}
\ee
where the function $h=h(y)$ is given {\it a priori}. If
coordinates $y^a$ are dimensionless then $h$ has the
dimension of moles. The density $h$ transforms simply
 when we pass from one
coordinate system in $Y$ to another: it is multiplied by
the Jacobian of the transformation. Of course, it is
always possible to change coordinates in such a
way that the numerical value of $h$ equals 1. Such
coordinates will be called {\it unimodular} and we will always use
them for the sake of simplicity. Such a formulation carries
however a danger: some expressions have to be multiplied by $h$
which is equal to {\it one mole}. This is very easy to
forget. In order to have $h$ equal to dimensionless {\it one} we
will always choose dimension $\sqrt[3]{mole}$ for the material
space coordinates $\wek ya$. This way the factor $h$ can really be
forgotten.

To describe the configuration of the material at a given instant
of time $t$  we assign the position in the physical space $X$ to
each particle. This means that we have a diffeomorphism

\begin{equation}
F:Y\longmapsto X  \label{konf}
\end{equation}
describing the configuration of the material at the time $t$. The
dynamical history of the material is given by the one parameter
family ${\cal F}$ of configurations:
$$
{\cal F} :Y\times U\longmapsto X \ .
$$
 If the physical space is parameterized by space coordinates $\wek
xk$ having the dimension of {\it length},
then the history can be described in terms of 3 functions
depending on four parameters:
$$
x^k=x^k(t,y^a)=x^k(\wek y{\alpha}) \ .
$$
These are field variables of the theory.
The laws of hydrodynamics
will be formulated in terms of the second order partial differential
equations for the fields. For this purpose we need to express
hydrodynamic quantities (the velocity and the density of the
fluid) in terms of the first derivatives of the fields. We denote
$$
\ten xk{\alpha}:=\diff {x^k}{\wek y\alpha } \ .
$$
The velocity vector $v$ has components
\be
 v^k=\frac{\partial x^k}{\partial t}  = \dot{x}^k = \ten xk0 \ .
\label{vel}
\ee
To express the density $\rho$ of the fluid we calculate the
transport of the volume form $r$ from the material space to the
physical space. This means that in formula (\ref{r}) we have to
substitute $\wek ya$ as  functions of $\wek xk$, using the inverse
$F^{-1}:X\mapsto Y$ of the configuration (\ref{konf}). Denoting
$$
\ten yak = \diff {\wek ya}{\wek xk} \ ,
$$
where $\wek ya =\wek ya (\wek xk)$ is the coordinate expression
for the mapping $F^{-1}$, we get:
\be
(F^{-1})^\ast (hdy^1 \wedge dy^2 \wedge dy^3) =
h\det(\ten yak)dx^1 \wedge dx^2 \wedge dx^3 \ . \label{Fro}
\ee
We conclude that the matter density (moles per volume!) is given by
the determinant: 
\be
\rho = h\det(\ten yak) = \det(\ten yak) \ ,
\ee
where the matrix ($\ten yak$) is considered not as a primary
quantity, but as a nonlinear function of the matrix ($\ten xka$),
namely its inverse. The inverse of the density is equal to the
molar volume $V$ of the fluid:
\begin{equation}
 V=\frac {\det\ (\ten xka )}h = \det\ (\ten xka ) \  \label{V}
 \end{equation}
or, in terms of the volume form:
\be
F^\ast (dx^1 \wedge dx^2 \wedge dx^3 )=
Vdy^1 \wedge dy^2 \wedge dy^3 \label{F*V} \ .
\ee

The laws of hydrodynamics
will be derived from the variational formula with the
field lagrangian equal to the difference between the
kinetic and the potential energy: 
\be
{\cal L} =\frac 12 Mv^2 - e(V) \label{lagr}
\ee  
(if the coordinates $\wek ya$ are not unimodular then the above
expression has to be multiplied by $h$). The constant $M$ denotes
the molar mass of the fluid (it could also be $y$--dependent for a
nonhomogeneous fluid). The internal energy $e=e(V)$ is given for
the given fluid. It determines the properties of the fluid
according to the fundamental equation:

 \begin{equation}
 de(V)=-pdV \ ,  \label{pdV}
 \end{equation}
where $p$ denotes the pressure. Due to equations (\ref{vel}) and
(\ref{V}) the lagrangian density (\ref{lagr}) can be considered as
a function of field variables and their first derivatives together
with the independent variables of the theory: \be
{\cal L} ={\cal L}(\wek xk ,\ten xk\alpha ;\wek y\alpha) \ .
\ee

We will show that the 
Euler--Lagrange equations of the
above field theory
$$
\frac{\partial {\cal L}}{\partial \wek xk} -
\base y{\alpha}  
\frac{\partial {\cal L}}{\partial \ten xk{\alpha}}= 0
$$
are equivalent to the equations of hydrodynamics. It is
useful for this purpose to analyse the canonical
structure of our theory. Introduce the momenta canonically
conjugate to field variables:

\begin{equation}
\ten {\cal P}{\alpha}k := \frac{\partial {\cal L}}{\partial \ten
xk{\alpha}}\ . \label{momenta} 
\end{equation}
This way the Euler--Lagrange equations generated by the lagrangian
$\cal L$ can be written as follows:

\begin{equation}
\base y{\alpha} \ten {\cal P}{\alpha}k = \diff {\cal L}{\wek xk} \
. \label{euler}
\end{equation}
Equations (\ref{momenta}) together with (\ref{euler}) can be
interpreted as the generating formula for the symplectic relation
(see \cite{Tulcz} and \cite{Kij-Tulcz}):

\begin{equation} 
\delta {\cal L} = \base y{\alpha}(\ten {\cal P}{\alpha}k 
\delta \wek xk ) = 
(\base y{\alpha}\ten {\cal P}{\alpha}k) \delta \wek xk +
\ten {\cal P}{\alpha}k \delta \ten xk{\alpha} \ . \label{eq:E-L}
\end{equation}
Using the explicit form (\ref{lagr}) of the lagrangian function and the
fundamental equation (\ref{pdV}) we obtain the following expression
for the canonical momenta:

\begin{equation}
\ten {\cal P}0k = M\kowek vk    
\end{equation}
and therefore $\ten {\cal P}0k$ represents the kinetic momentum 
density of the fluid,

\begin{equation}  
\ten {\cal P}ak = p \diff V{\ten xka} = pV\ten yak \label{eq:Pa}
\ .  \end{equation}
To prove the
equation (\ref{eq:Pa}) we used the fact that the derivative of the
determinant with respect to a matrix element is equal to the
corresponding element of the inverse matrix times the determinant
itself.

The Euler--Lagrange equations (\ref{euler}) can thus be written as
follows: 

\begin{equation}
M\dot v_k = -\base ya (pV\ten yak) = -(\ten yak \diff p{\wek ya})V-
p\base ya (V\ten yak) \ .  \label{r.Eulera}
\end{equation}
Observe that 
$$
\ten yak \diff p{\wek ya} = \diff {\wek ya}{\wek xk}\diff p{\wek
ya}= \diff p{\wek xk} \ .
$$
Moreover, we prove that the last term of (\ref{r.Eulera}) vanishes
identically. Indeed, due to (\ref{F*V}) we have:
 $$
0 = F^\ast d(\base xk \rfloor dx^1 \wedge dx^2 \wedge dx^3) = 
d[(\ten yak \base ya) \rfloor F^\ast (dx^1 \wedge dx^2 \wedge
dx^3)]=
 $$ 
\be
= d[(V\ten yak )\base ya \rfloor dy^1 \wedge dy^2 \wedge dy^3]=
\base ya (V\ten yak ) dy^1 \wedge dy^2 \wedge dy^3 \ .
\label{dVyak}
  \ee 
Equation (\ref{r.Eulera}) takes therefore the form of
the Newton equation: 
\be
M \rho \dot v_k =-\diff p{\wek xk} \ , \label{r.Newtona}
\ee
where the force on the right--hand side is given by the gradient of
the pressure.

The dynamical equations of the theory imply the energy and momentum
conservation laws via the so called N\"{o}ther theorem. We
introduce for this purpose the following energy--momentum tensor:

\begin{equation}
\ten {\cal T}\alpha \beta =\ten {\cal P}\alpha k \ten xk\beta
-\delta ^\alpha _\beta {\cal L}  \label{tenenp} \ .
\end{equation}
To prove the N\"{o}ther theorem we observe that

\begin{equation}
\base y\beta {\cal L}(x^k,{\ten xk\alpha };{y^\alpha })=
\diff{\cal L}{x^k} \ten xk\beta +
\diff{\cal L}{\ten xk\alpha }x^k_{\alpha \beta} + 
\diff{\cal L}{y^\beta } \ , \label{dL}
\end{equation}
where by $x^k_{\alpha \beta}$ we denote the second derivatives:
$$
x^k_{\alpha \beta}=\base x\beta (\ten xk\alpha ) = 
\frac{\partial ^2\wek xk}{\partial \wek x\beta \partial \wek
x\alpha} \ . 
$$
Combining (\ref{tenenp}), (\ref{dL}) and the definition 
(\ref{momenta}) of the
canonical momenta we obtain:

\begin{equation}
\base y\beta \ten {\cal T}{\beta}{\alpha} + 
\diff{\cal L}{y^\alpha }=
(\base y\beta \ten {\cal P}{\beta}k - \diff{\cal L}{x^k}) \ten
xk\alpha  +  (\ten {\cal P}{\beta}k-\diff{\cal L}{\ten xk\beta } )
x^k_{\alpha \beta} = 
(\base y\beta \ten {\cal P}{\beta}k - \diff{\cal L}{x^k}) \ten
xk\alpha
\ . \label{Noet.} 
\end{equation}
Therefore the Euler--Lagrange 
equations (\ref{euler}) imply:  
\begin{equation} 
\base y\beta \ten {\cal T}{\beta}{\alpha} +\diff{\cal L}{y^\alpha
}=0 \label{divT} \ . 
\end{equation} 
The above equality is called in the field
theory  {\it N\"{o}ther theorem} (see \cite{Noether}). We
notice that from 3 equations of motion (\ref{euler}) we get 4
equations (\ref{divT}). This means that they are not independent.
Indeed equations (\ref{momenta}) and (\ref{Noet.}) imply:  
$$
(\diff{\cal L}{y^a} + \base y\beta \ten {\cal T}{\beta}a )\ten
yak \wek vk=(\base y\beta \ten {\cal P}{\beta}l - \diff{\cal
L}{x^l}) \ten xla \ten yak \ten xk0= (\base y\beta \ten {\cal
P}{\beta}l - \diff{\cal L}{x^l})\ten xl0=$$
$$
=\diff{\cal L}{t}+\base y\beta \ten {\cal T}{\beta}0 \ ,
$$
so the  components of the N\"other theorem corresponding to 
$a=1,2,3$  imply its
0--component. Due to the equality (\ref{Noet.}) they are
equivalent with the field equations.

Using the explicit form of the lagrangian $\cal L$ we
obtain the following expressions for the components
of the energy--momentum tensor:

\begin{equation}
\ten {\cal T}00 = \frac{1}{2} Mv^2 + e
\end{equation}
and therefore $\ten {\cal T}00$ represents the total energy density,

\begin{equation}
\ten {\cal T}a0 = pV \ten yak v^k \ ,
\end{equation}

\begin{equation}
\ten {\cal T}0a = Mv_k \ten xka \ ,
\end{equation}

\begin{equation}
\ten {\cal T}ab = \delta ^a_b (e + pV - \frac{1}{2} Mv^2) \ .
\end{equation}
Especially interesting is the case of the lagrangians which do not
depend on the independent parameters of the theory. In the case
of hydrodynamics this happens e.g. when the fluid is homogeneous
(both the mass and the state equation are the same at different
points of the material). In this case the lagrangian depends only
on $\wek xk$ and $\ten xk{\alpha}$ and we obtain:
\be
\base y\beta \ten {\cal T}{\beta}{\alpha} =0 \label{divT=0} \ . 
\ee
 The
0--th component of the equation (\ref{divT=0}) expresses the energy
conservation:

\begin{equation}
\diff{}{t} \ten {\cal T}00 + \base ya \ten {\cal T}a0 =\diff{}{t}(
\frac{1}{2} Mv^2 + e) + \base ya (pV \ten yak v^k)=0 \ . 
\end{equation}
Integrating the above equality over a finite domain $D$ and using
the Stokes theorem we can convert the volume integral of the last
term into the surface integral over the boundary $\partial D$ of
$D$. The integral is equal to  (minus) the work
performed by the pressure on the boundary of $D$. The work
performed at the boundary is the only reason for the change of
the  total energy of the fluid
contained in $D$. Using the identity (\ref{dVyak})  we can also
rewrite the last term in the above equality as follows:
$$ \base ya (pV \ten yak v^k)=V\base xk (pv^k) \ . $$ 
Finally, we get the energy
conservation law in the standard form: 
$$
\diff{}{t}(\frac{1}{2} Mv^2 + e) + V\base xk (pv^k) =0 \ . 
$$ 
Using (\ref{pdV}) the ``material'' components of the equation
(\ref{divT}) take the following form: 
$$   
\diff{}{t} \ten {\cal T}0b + \base ya \ten
{\cal T}ab = \diff{}{t}(Mv_k \ten xkb) + \base ya [\delta ^a_b (e +
pV - \frac{1}{2}Mv^2)]=
$$ 
\begin{equation}
= \ten xkb[M\diff{v_k}{t}+\base xk (e + pV)]=
\ten xkb(M\diff{v_k}{t}+ V\base xk p) = 0 \ . 
\end{equation} 
The last equality is equivalent to (\ref{r.Newtona}) and represents
the momentum conservation law of the perfect fluid.

\section{Euler picture}
In the previous section we described the configuration of
a continuous medium as a mapping $F$ from its material space $Y$ to
the physical space $X$. Now we invert the role of the independent
parameters and the field variables of our theory. To describe the
same configuration of the material we will use the inverse mapping
$G=F^{-1}$: $$
G: X\longmapsto Y \ .
$$
The time coordinate $x^0=t$ together with 
coordinates $(x^k)$ form a system  of spacetime coordinates $(x^{\mu})$ ($\mu =
0,1,2,3$) in spacetime $M:= U\times X$. Now the dynamical history
of the material is given by the one parameter family $\cal G$ of
configurations: 
$$ 
{\cal G} : M \longmapsto Y \ . 
$$
The laws of the hydrodynamics can be expressed as  second
order partial differential equations for the field variables
$\wek ya (\wek x\mu)$ which are the functions on the
spacetime $M$. For this purpose we express hydrodynamic
quantities (the velocity and the density of the fluid) in terms of
the first derivatives of the fields. We already know the expression for the
density:

\be  
\rho=\det (\ten yak) \ . \label{ro}
\ee 
Moreover,
$$
0=\frac d{dt} x^k(t, y^a(t, x^l)) = 
\diff {x^k}t (t, y^a(t, x^l)) +
\ten xka (t, y^a(t, x^l))\ten ya0 (t, x^l) \ ,
$$
which implies:

\be
\wek vk= \diff {x^k}{t} =-\ten xka {\dot y}^a \ . \label{vv}
\ee
The above formula expresses the velocity $v^k$ in terms of the
derivatives of the eulerian field variables $y^a $ if we consider
the matrix $(\ten xka )$ not as a primary quantity, like in the
lagrange picture, but as a nonlinear function of $(\ten yak )$,
namely its inverse.

It is
important to notice that the above quantities satisfy {\it
automatically} the continuity equation and therefore we do not need
to postulate it independently. The situation is similar as in the
classical electrodynamics, where the fields which have been derived
from the electromagnetic vector--potential satisfy  automatically
the first pair of the Maxwell equations. The second pair becomes the
dynamical 2--order equations for the potentials. We can say that
our field variables $y^a$ play the role of the ``potentials'' for
the hydrodynamic quantities ($v^k,\rho$). The analogy with the
electrodynamics goes further: there are gauge transformations in
both theories which do not change the physical quantities. In the
electrodyna\-mics gauge transformations consist in adding a
gradient to the electromagnetic vector--potential. Here, gauge transformations are
given by unimodular reparameterizations of the material space $Y$.
Such a reparameterization consists  only in ``changing
names to the particles of the material'' and does not change the
values of $v^k$ and $\rho$.
  
To prove the 
continuity equation we first introduce the matter current $j$ 
as a vector density defined  by the formula:
\be
j^\kappa=\epsilon^{\kappa \mu \nu \lambda}\ten y1\mu \ten y2\nu \ten
y3\lambda \ . \label{defj}
\ee
We have 
$$
j^0=\epsilon^{0 \mu \nu \lambda}\ten y1\mu \ten y2\nu \ten
y3\lambda= 
\epsilon^{kmn}\ten y1k \ten y2m \ten y3n=
\det (\ten yak ) =\rho \ .
$$ 
Moreover, we have:
$$
j^k \ten yak =\epsilon^{k\mu \nu \lambda}\ten y1\mu \ten y2\nu
\ten y3\lambda \ten yak = - \epsilon^{0 \mu \nu \lambda}\ten y1\mu
\ten y2\nu \ten y3\lambda \ten ya0 +
\epsilon^{\kappa \mu \nu \lambda}\ten y1\mu \ten y2\nu \ten
y3\lambda \ten ya\kappa \ .
 $$ 
The last term vanishes being the determinant of the $4 \times 4$
matrix which has necessarily 2 columns equal. Therefore
$$
j^k \ten yak= -\rho \ten ya0 \ ,
$$
which finally implies:
$$
j^k = -\rho \ten ya0 \ten xka = \rho v^k \ .
$$
Introducing now the ``four--velocity'' $u^\mu$ defined as follows: 
$$
u^0:=1 \ ,
$$ 
$$
 u^k:=v^k \ ,
$$ 
we have
$$
j^\mu = \rho u^\mu \ .
$$

We will now show that the continuity equation for the matter
current $j$ is a consequence of the fact that the volume form $r$ 
is closed since its external differential necessarily vanishes
being a 4--form in a 3--dimensional space $Y$.  For this purpose we
prove that the transport of $r$ to the space--time $M$ (such a
transport is a differential 3--form in the 4--dimensional manifold
i.e. the vector--density in $M$) is equal to $j$. Indeed:

$$
{\cal G}^\ast (r) =
{\cal G}^\ast (dy^1 \wedge dy^2 \wedge dy^3) = \ten y1\mu \ten
y2\nu \ten y3\lambda dx^\mu \wedge dx^\nu \wedge dx^\lambda \ .  
$$
Using the identity:
$$
dx^\mu \wedge dx^\nu \wedge dx^\lambda = 
\epsilon^{\kappa \mu \nu \lambda}\base x\kappa \rfloor  
dx^0 \wedge dx^1 \wedge dx^2 \wedge dx^3
$$ 
 and definition (\ref{defj}) we get
\be
{\cal G}^\ast (r) = j^\kappa \base x\kappa \rfloor  
dx^0 \wedge dx^1 \wedge dx^2 \wedge dx^3 \label{G*r} \ .
\ee
Finally we get the continuity equation as a consequence of (\ref{G*r})
and the fact that the 3--form $r$ is closed: 
\be 
0={\cal G}^\ast (dr) = d\left(
j^\mu \base x\mu \rfloor  dx^0 \wedge dx^1 \wedge dx^2 \wedge dx^3 \right)
= \diff {j^\mu}{ x^\mu} 
dx^0 \wedge dx^1 \wedge dx^2 \wedge dx^3 \ . \label{dj=0}
\ee
Observe that replacing $r$ by $Mr$ we obtain the same result for
the current $Mj$. This means that the following identity is valid:
$$
\base x\mu (M\rho u^\mu ) =0
$$
even for the non--homogeneous fluid.

 Similarly as in the Lagrange picture, the field equations can be
derived from the variational principle. Obviously, the numerical
value of the action integral corresponding to a given configuration
has to be equal in both the Lagrange picture and the Euler picture.
This means that the new lagrangian $L$ and the old lagrangian
${\cal L}$ are related via the following formula:
 $$ 
L dx^1 \wedge dx^2 \wedge dx^3= G^{\ast} ({\cal L}dy^1 \wedge
dy^2 \wedge dy^3) = 
\rho {\cal L} dx^1 \wedge dx^2 \wedge dx^3 \ .
$$ 
We conclude that the numerical value of the new lagrangian equals:
\be  
L=\rho (\frac 12 M\wek v2 - e) \ ,\label{Elagr}
\ee 
but now we have to express it in terms of the field variables 
$\wek ya $ and
their derivatives $\ten ya{\mu} $ using equations (\ref{ro}) and
(\ref{vv}). Again, the Euler--Lagrange equations together
with the definition of the momenta $\ten {\Theta}{\mu}a$ 
canonically
conjugate to field  variables $y^a$ can be
interpreted as a generating formula for the symplectic relation:

\begin{equation} 
\delta L (\wek ya,\ten ya{\mu};x^{\mu}) = \base x{\mu}(\ten
{\Theta}{\mu}a \delta \wek ya ) =  (\base x{\mu}\ten
{\Theta}{\mu}a) \delta \wek ya + \ten {\Theta}{\mu}a \delta \ten
ya{\mu}   
\end{equation}
or, equivalently:
\be
\ten {\Theta}{\mu}a = \diff {L}{\ten ya\mu }
\ee
and

\be
\base x\mu\ten {\Theta}{\mu}a = \diff {L}{y^a} \ .
\label{r.EL}
\ee

The explicit form
(\ref{Elagr}) of the lagrangian $L$ implies the following
expressions for the components of $\ten {\Theta}{\mu}a$. Using the
fundamental equation (\ref{pdV}) which -- in terms of the density
$\rho$ -- reads 
$$
de=\frac p{\rho ^2} d\rho \ ,
$$
we obtain:
$$
\ten {\Theta}0a = -\rho \ten xka M\kowek vk \ , 
$$
$$
\ten {\Theta}ka = -\rho \ten xla [M\kowek vl \wek vk +
\delta ^k_l (e + pV - \frac 12 M\wek v2)] \ .
$$
Therefore, the field equations can be written as follows:
$$
\base x\mu \ten {\Theta}{\mu}a = -\base x\mu
(\rho \ten xka M\kowek vk u^\mu ) -
\base xk [\rho \ten xka (e + pV - \frac 12 M\wek v2) =
$$
\be=
-\ten xka [\diff{p}{x^k} + \base x\mu (\rho u^\mu M\kowek vk )]=0
\label{r.pola} \ .
\ee
Since the matrix $\ten xka$ is non--degenerate, the equations can
be rewritten in the following, equivalent form:
\be
0=\diff{p}{x^k} + \base x\mu (\rho u^\mu M\kowek vk )=
\diff{p}{x^k} + M\rho u^\mu \base x\mu \kowek vk \ ,
\label{r.pola2}
\ee
where we have used the continuity equation. The operator 
$$
\frac d{dt} := u^\mu \base x\mu 
$$
is called the ``substantial'' derivative. It is obviously
equivalent to the partial time derivative in the Lagrange picture
(i.e. with $\wek ya$ being constant). The equations 
(\ref{r.pola2}) are called Euler equations. They are equivalent to
the field equations (\ref{r.Newtona})
in the Lagrange picture.

As in any other field theory we can also prove
the N\"other theorem
for the following energy--momentum tensor:
\be
\ten t{\mu}\nu = \ten {\Theta}{\mu}a \ten ya\nu - 
\delta ^\mu _\nu L
\ .
\ee
The theorem consists in deriving the equivalence between field
equations and the (non--)conservation laws. Its proof is completely
analogous to the one given in the previous section.
Simple calculations lead to the identity:

\be
\base x\mu \ten t{\mu}{\nu} + \diff {L}{x^\nu} = 
(\base x\mu\ten {\Theta}{\mu}a - \diff {L}{y^a})\ten ya\nu
\label{dt+L} \ .
\ee
Therefore, the field equations (\ref{r.EL}) imply the 
(non--)conservation
laws:
\be
\base x\mu \ten t{\mu}{\nu} + \diff {L}{x^\nu} = 0 \label{divt}
 \ .
\ee
 Similarly as in the Lagrange picture only 3 among them are
independent (this is due to the fact that the continuity
equation has been incorporated {\it a priori} into the
structure of the theory). Moreover, they are completely equivalent
to field equations (\ref{r.pola}). Indeed, equation (\ref{dt+L})
implies the following identity:

$$
-(\base x\mu \ten t{\mu}k + \diff {L}{x^k})\wek vk = 
(\base x\mu\ten {\Theta}{\mu}a - \diff {L}{y^a})\ten yak 
\ten xkb \ten yb0 =
$$
\be 
=(\base x\mu\ten {\Theta}{\mu}a - 
\diff {L}{y^a}) \ten ya0 =
\base x\mu \ten t{\mu}0 + \diff {L}{x^0} \ .
\ee

Again, the symmetry of the lagrangian with 
respect
to the space--time translations:
$$
\base x\nu L =0 \ ,
$$
which is physically equivalent to the absence of the external
forces, implies the conservation laws:
\be
\base x\mu \ten t{\mu}{\nu}  = 0 \ .
\ee
Simple calculations lead to the following formulae for the
components of the energy--momentum tensor $\ten t{\mu}\nu$ derived
from the lagrangian $L$:
$$ 
\ten t00 = \rho (\frac 12 M\wek v2 + e)
$$
$$
\ten tk0 = \rho \wek vk (\frac 12 M\wek v2 + e + pV)
$$
$$
\ten t0k = -\rho M\wek vk
$$
$$
\ten tkl = - p\delta ^k_l - \rho M \wek vk \kowek vl \ .
$$
Equation $\partial _\mu \ten t{\mu}0 = 0$ represents therefore the
energy conservation law and equation $\partial _\mu \ten t{\mu}k = 0$
expresses the momentum conservation law.

The following identities express relations
between the objects used in the Lagrange and Euler pictures:
$$
\ten {\Theta}0a = -\rho \ten {\cal T}0a 
$$
$$
\ten {\Theta}ka = -\rho \wek vk \ten {\cal T}0a   - 
\rho \ten xkb \ten {\cal T}ba
$$
$$
\ten {\cal P}0k = -V\ten t0k   
$$
$$  
\ten {\cal P}ak =V{\ten yal} \wek vl \ten t0k - V \ten yal \ten
tlk  \ .  
$$
We see that in the rest frame (i.e. when $v^k=0$) the canonical
momentum in one picture corresponds (up to a sign) to the
energy--momentum tensor in the other picture and vice versa.

\section{Thermodynamics}

To describe thermal properties of the fluid we need one more
potential (see \cite{Kij-Tulcz2} and \cite{KSG}).
As we will see in the sequel, the new potential $\tau$ can be
treated as a "material time". We will describe the thermo--dynamical
history of an isoentropic flow in terms of 4 field potentials
depending on 4 independent variables. We already know that there
are 2 possible choices as far as the ``spatial'' coordinates are
considered: the Lagrange and the Euler picture. Similarly, in
both cases we may choose the physical time $t$ or the ``material
time'' $\tau$ as the independent variable and the remaining
parameter as the field potential. This way we obtain 4 possible
pictures of the thermo--hydrodynamics. In the present paper we
are going to discuss only two of them, keeping always the
physical time together with the physical space $X$ and the
``material time'' together with the material space $Y$.
Remaining two pictures have no natural relativistic counterpart
and are less interesting, although technically it is very easy
to formulate them.

In the present section we start with the complete Euler picture,
based on the choice of the physical space--time $M$ as the space
of independent parameters. To describe the space of field
potentials we add  new dimension $\tau$ to the matter
space $Y$. This way we obtain the 4--dimensional matter spacetime
$Z$ with $\tau$ playing the role of a "material time" (having {\it a
priori} nothing to do with the physical time $t=x^0$). Let us  
introduce coordinates $\wek z\alpha $ ($ \alpha = 0,1,2,3$)
in the material spacetime $Z$ putting $z^0 =\tau $ and $z^a =y^a$. 
We
will describe the history of the fluid in
terms of 4 potentials $z^\alpha =z^\alpha (x^\mu )$. The potentials
 give the coordinate expression for the mapping:
$$
{\cal G} : M \longmapsto Z \ .
$$
 The laws of
the thermo--hydrodynamics will be formulated as a system of
2--order partial differential equations for the potentials. For this
purpose we have to express the thermo--dynamical quantities in
terms of the derivatives of the potentials. We already know how to
do it for the hydrodynamical quantities $j^\mu $. For the
temperature $T$ we choose
the following ansatz:

\begin{equation}
T = \beta (\dot {\tau} + \wek vk \diff{\tau}{\wek xk})  = 
\beta \wek u\mu \ten z0\mu   \label{deftemp} 
\end{equation}
with $\beta $ being a positive phenomenological constant.
The
microscopic interpretation of the above definition will
be given in the sequel but in the present section we adopt a purely
phenomenological point of view where $\tau $ is merely a potential
for the temperature, similarly as $y^a $ were the potentials for the 
hydrodynamic quantities $j^\mu $. We will prove that the choice of
the Helmholtz free energy $f(V,T)$ as a potential part of
the Lagrangian $L$, i.e.
\be
	L =\frac 1V \left[ \frac 12 Mv^2 - f(\frac 1V,T) \right] 
=\rho \left[ \frac 12 Mv^2 - f(\rho,T) \right] \label{Lf}	
\ee
leads to the field equations which are equivalent with the laws of
the ther\-mo--hy\-drodyna\-mics of isoentropic flows.

Again, the Euler--Lagrange equations together with the definition of
the momenta $\ten {\Theta}{\mu}{\alpha}$ canonically conjugate to the 
field  variables $z^\alpha$ can be
interpreted as a generating formula for the symplectic relation:

\begin{equation} 
\delta L(\wek z\alpha ,\ten z{\alpha}{\mu} ; x^\mu )
= \base x{\mu}(\ten {\Theta}{\mu}\alpha \delta \wek z\alpha ) = 
(\base x{\mu}\ten {\Theta}{\mu}\alpha) \delta \wek z\alpha +
\ten {\Theta}{\mu}\alpha \delta \ten z{\alpha}{\mu} \label{generL}
\end{equation}
or, equivalently:
\be
\ten {\Theta}{\mu}\alpha = \diff {L}{\ten z{\alpha}\mu }
\ee
and

\be
\base x\mu\ten {\Theta}{\mu}\alpha = \diff {L}{z^\alpha}
\label{r.EL4} \ .
\ee

The explicit form
(\ref{Lf}) of the lagrangian $L$ enables us to calculate the
components of $\ten {\Theta}{\mu}\alpha$ in terms of the field
potentials and their derivatives. Using the fundamental equation: 
\be df=-pdV -sdT =
\frac p{\rho ^2} d\rho - sdT \label{sdT}
\ee
with $s$ being the molar entropy of the fluid, we obtain:
\be
\ten {\Theta}{\mu}0 = \beta \rho s u^\mu \ , \label{tetamu0}
\ee
\be
\ten {\Theta}0a = -\rho \ten xka (M\kowek vk + \beta s\ten z0k )
\ , \label{teta0a}  
\ee
$$
\ten {\Theta}ka = -\rho \ten xla [(M\kowek vl + \beta s\ten z0l )
\wek vk + \delta ^k_l (f  - \frac 12 M\wek v2 + pV)] \ .
$$

 The 0--th component of field equations gives thus 
the entropy conservation law:
\be
 \base x\mu \ten {\Theta}{\mu}0 = \beta \base x\mu (\rho s u^\mu) 
= \beta \rho u^\mu \base x\mu  s = 0 \ . \label{konsS}
\ee 
The remaining equations read:
$$
\base x\mu \ten {\Theta}{\mu}a = -\base x\mu
[\rho \ten xka (M\kowek vk +\beta s\ten z0k ) u^\mu ] -
\base xk [\rho \ten xka (f + pV - \frac 12 M\wek v2) =
$$
\be=
-\ten xka [\diff{p}{x^k} + \base x\mu (\rho u^\mu M\kowek vk )]
-\beta \ten xka \ten z0k \base x\mu (\rho s u^\mu) =0
\label{r.pola4} \ .
\ee
The last term vanishes due to the entropy conservation and the
above equations become simply the Euler equations discussed
 in the previous sections. 

Similarly as in any field theory we obtain the N\"other theorem
relating the translational symmetries of the system with the energy
and momentum conservation laws. The 
energy--momentum tensor of the thermo--hydrodynamic system is
given by the standard formula: 
\begin{equation} 
\ten t\mu \lambda =
\ten {\Theta}{\mu}\alpha \ten z\alpha \lambda -\delta ^\mu
_\lambda L \ . \label{tmunu} 
\end{equation} 
Using the explicit expressions for canonical momenta $\ten
{\Theta}{\mu}\alpha$ we obtain:
$$ 
\ten {\Theta}0a \ten ya0 = \rho(M\wek v2 + \beta s\kowek vk \ten
z0k ) $$
$$
\ten {\Theta}0a \ten yak = -\rho(M\kowek vk + \beta s \ten z0k )
$$
$$ 
\ten {\Theta}ka \ten ya0 = \rho v^k(\frac 12 M\wek v2 + \beta
s\kowek vl \ten z0l + f + pV )
$$
$$
\ten {\Theta}ka \ten yal = -\rho [(M\kowek vl + \beta s\ten z0l )
\wek vk + \delta ^k_l (f + pV - \frac 12 M\wek v2)] \ . 
$$
 Moreover
$$
\ten {\Theta}{\mu}0 \ten z0\lambda = \beta \rho s u^\mu\ten
z0\lambda \ .	
$$
Finally, we obtain the formula for $\ten t\mu \lambda$:
$$ 
\ten t00 = \rho (\frac 12 M\wek v2 + e)
$$
$$
\ten tk0 = \rho \wek vk (\frac 12 M\wek v2 + e + pV)
$$
$$
\ten t0k = -\rho M\wek vk
$$
$$
\ten tkl = - p\delta ^k_l - \rho M \wek vk \kowek vl
$$
with internal energy defined by 
$$
 e:=f+Ts \ .
$$ 

As in any other field theory we can also prove
the N\"other theorem
for the energy--momentum tensor $\ten t\mu \lambda$.
The theorem consists in deriving the equivalence between the field
equations and the (non--)conservation laws. Its proof is completely
analogous to the one given in the previous sections.
Simple calculations lead to the identity:

\be
\base x\mu \ten t{\mu}{\nu} + \diff {L}{x^\nu} = 
(\base x\mu\ten {\Theta}{\mu}\alpha - \diff {L}{y^\alpha})
\ten z{\alpha}\nu
\label{4dt+L} \ .
\ee
Therefore, the field equations (\ref{r.EL4}) imply the 
(non--)conservation
laws:
\be
\base x\mu \ten t{\mu}{\nu} + \diff {L}{x^\nu} = 0 \label{4divt}
\ee
 and vice versa because matrix $\ten z{\alpha}\nu$ is invertible.

We stress that now the 4 conservation laws are independent and
equivalent to the 4 field equations. In particular, the
energy--momentum conservation laws imply -- in our formulation --
the entropy conservation. This is due to the fact that the latter
is incorporated {\it implicite} in the structure of the theory via
the fact that the Lagrangian $L$ does not depend on $z^0$.

Again, the symmetry of the lagrangian with 
respect
to the space--time translations:
$$
\base x\nu L =0 \ ,
$$
which is physically equivalent to the absence of the external
forces, implies the energy--momentum conservation laws:
 
\begin{equation}
\base x\mu \ten t\mu \lambda = 0 \    \label{konsT} \ .
\end{equation} 
 
In the
above formulation the phenomenological constant $\beta$ may be
chosen arbitrarily. The choice  $\beta =1$ is also possible. It
gives us [time\ $\times$\ temperature] for the dimension of the new
potential $\tau=z^0$. As we shall see in the sequel, it is 
more natural to choose the dimension of $\beta$ equal to the
temperature and to measure $\tau$ in units of time.

\section{Hamiltonian formulation of thermo--hy\-dro\-dy\-namics}

The hamiltonian description of the thermo--hydrodynamics can be
obtained by the following standard, field--theoretical, Legendre
transformation (see e.g. \cite{Noether} or \cite{Kij-Tulcz}). The
generating formula (\ref{generL})  can be rewritten as follows:
$$
\delta L
= \base x{\mu}(\ten {\Theta}{\mu}\alpha \delta \wek z\alpha ) =
\base x0(\ten {\Theta}0\alpha \delta \wek z\alpha ) +
\base xk(\ten {\Theta}k\alpha \delta \wek z\alpha ) =
$$
$$
= {\ten {\dot \Theta}0\alpha} \delta \wek z\alpha +
\ten {\Theta}0\alpha \delta {\wek {\dot z}\alpha} + 
\base xk(\ten {\Theta}k\alpha \delta \wek z\alpha ) =
$$
\be
=\delta ({\ten {\Theta}0\alpha} {\wek {\dot z}\alpha}) -
 {\wek {\dot z}\alpha} \delta {\ten {\Theta}0\alpha}   +
 {\ten {\dot \Theta}0\alpha } \delta {\wek z\alpha} +
\base xk(\ten {\Theta}k\alpha \delta \wek z\alpha ) \ .
\end{equation}
Introducing the hamiltonian density:
$$
 H = \ten {\Theta}0\alpha { \wek {\dot z}\alpha } - L = \ten
t00  $$
we get another generating formula:
$$
-\delta  H = {\ten {\dot \Theta}0\alpha } \delta {\wek z\alpha}
- {\wek {\dot z}\alpha} \delta \ten {\Theta}0\alpha + 
\base xk(\ten {\Theta}k\alpha \delta \wek z\alpha ) \ ,
$$
where, at given time $t$, the function $H$ has to be expressed in
terms of the ``hamiltonian variables'' (or ``canonical
variables''), i.e. in terms of the fields $\wek z\alpha$ and of the
time--like component $\ten {\Theta}0\alpha$ of the momenta. It is
useful to choose the special notation for the latter: 
$$ 
\pi _\alpha := \ten {\Theta}0\alpha \ . 
$$
We have therefore:
\be
-\delta H 
= \kowek {\dot \pi}\alpha  \delta {\wek z\alpha}
- \wek {\dot z}\alpha \delta \kowek {\pi}\alpha + 
\base xk(\ten {\Theta}k\alpha \delta \wek z\alpha ) \ .
\label{generH}
\ee
As we already know, the numerical value of our
hamiltonian is simply: 
$$
H = \rho (\frac 12 M\wek v2 + e) \ ,
$$
where now we consider the internal energy $e$ as a function of
$\rho $ and $s$. It contains the entire information about the
thermo--hydrodynamic properties of the fluid due to the
corresponding fundamental equation:
\be
de=-pdV+Tds=\frac p{\rho ^2} d\rho +Tds \ . \label{de=}
\ee
In order to complete the Legendre transformation
we observe that the formulae (\ref{tetamu0})
and (\ref{teta0a}) imply:  
\be 
s = \frac {\kowek {\pi}0}{\beta \rho} \ ,\ \ \ \ \ \ \ \ \ \ \ \ 
\label{sofpi}
\ee
\be
p_k:= Mv_k = - \frac {\kowek {\pi}\alpha \ten z{\alpha}k}{\rho } \ .
\label{vkofpi}
\ee
Together with the formula (\ref{ro}), they allow us to express the 
hamiltonian density $H$ in terms of canonical variables 
$(\wek z{\alpha} ,\kowek {\pi}\alpha )$ and their space--like
derivatives. The formula (\ref{generH}) 
can be rewritten as follows:
\be
-\delta H(\wek z\alpha ,\ten z{\alpha }k,\pi _\alpha ) 
= (\kowek {\dot \pi}\alpha  + \base xk \ten {\Theta}k\alpha )
\delta {\wek z\alpha} +
\ten {\Theta}k\alpha \delta \ten z{\alpha }k  
- \wek {\dot z}\alpha \delta \kowek {\pi}\alpha  \ ,
\label{generHinf}
\ee
which is obviously equivalent to the dynamical equations:
$$
\ten {\Theta}k\alpha = - \diff H{\ten z{\alpha }k} \ ,
$$
$$
\kowek {\dot \pi}\alpha  + \base xk \ten {\Theta}k\alpha =
- \diff H{\wek z\alpha} 
$$
and
$$
\wek {\dot z}\alpha =  \diff H{\kowek {\pi}\alpha } \ .
$$
The first two equations are usually written with the help of the
variational derivative:
\be
-\kowek {\dot \pi}\alpha =
\frac{\delta H}{\delta \wek z\alpha} :=
 \frac{\partial H}{\partial \wek z\alpha}
- \base xk \frac{\partial H}{\partial \ten z{\alpha }k}
\ .
\label{Hamilton2}
\ee
Also the last equation can be rewritten in terms of the
variational derivative since the hamiltonian does not depend on
the derivatives of $\kowek {\pi}\alpha$:
\be
\wek {\dot z}\alpha = 
\frac{\delta H}{\delta \kowek {\pi}\alpha} \ .
\label{Hamilton1}
\ee

The theory can be also formulated in the
language of infinite--dimensional hamiltonian systems (see e.g.
\cite{Chernoff-Marsden}). The infinite--dimensional
phase space of the system is a
functional space of 8 functions $(\wek z{\alpha} ,\kowek
{\pi}\alpha )$ defined on a fixed domain $D\subset X$ in the
physical space $X$  (of
course, $D$ can also be equal $X$ if we want to describe the world
filled entirely with the fluid). The dynamics of the system is
governed by the hamiltonian $H_D$ equal to the integral
of the hamiltonian density $H$:
$$
H_D := \int _D H(\wek z{\alpha} (x) ,\kowek {\pi}\alpha (x))
d^3x \ ,
$$
and the generating formula (\ref{generH}) becomes now:
\be
-\delta H_D = 
\int _D \kowek {\dot \pi}\alpha  \delta {\wek z\alpha}
- \wek {\dot z}\alpha \delta \kowek {\pi}\alpha +
\int _{\partial D} 
\ten {\Theta}{\perp}\alpha \delta \wek z\alpha  \ , 
\label {generHcal}
\ee
where $\partial D$ denotes the boundary of $D$ and by 
$\ten {\Theta}{\perp}\alpha $ we mean the transversal (with respect
to $\partial D$)  component of the momentum ${\Theta}$. The above
definition of our infinite dimensional phase space is not
complete unless we specify some boundary conditions which enable
us to annihilate the boundary integral in the above formula in
order to obtain the infinite dimensional analog
\be
-\delta H_D = 
\int _D \kowek {\dot \pi}\alpha  \delta {\wek z\alpha}
- \wek {\dot z}\alpha \delta \kowek {\pi}\alpha 
\label{generHcal0}
\ee
of the finite dimensional generating formula:
$$
-dH(q,p)={\dot p}_a dq^a - {\dot q}^a dp_a \ .
$$
The simplest way to remove 
the boundary integral is to impose
the boundary conditions $z^\alpha \vert
_{\partial D}$ for the configuration variables. Physically, keeping
the values of $z^a \vert
_{\partial D}$ constant in time is equivalent to the
condition $v^k \vert
_{\partial D}=0$. Moreover, controlling the value of 
$z^0 \vert _{\partial D}$ (not necessarily constant in time) means
that we control the temperature on the boundary. This means that the
system is kept in a thermal bath. Mathematically, the above
boundary conditions mean that we consider the phase space {\bf P}
of functions $z^\alpha$  which fulfill the Dirichlet
conditions on $\partial D$. The hamiltonian dynamics
 in {\bf P} generated by the
hamiltonian $H_D$ describes the mixed Dirichlet--Cauchy
problem for the field equations. Within the space {\bf P} of
functions fulfilling the boundary conditions we have  $$ 
\delta
z^\alpha \vert_{\partial D} = 0 
$$
which enables us to integrate by parts 
during all the calculations and to neglect all the boundary integrals.

Of course, controlling the Dirichlet data is not the only way to
eliminate the boundary integrals. Using the identity
$$
\ten {\Theta}{\perp}\alpha \delta \wek z\alpha =
\delta (\ten {\Theta}{\perp}\alpha  \wek z\alpha ) -
\wek z\alpha \delta \ten {\Theta}{\perp}\alpha   
$$ 
and defining the new hamiltonian
$$
F_D:=H_D - \int _{\partial D} 
\ten {\Theta}{\perp}\alpha  \wek z\alpha 
$$
we can rearrange the formula (\ref{generHcal}) as follows:
\be
-\delta F_D = 
\int _D \kowek {\dot \pi}\alpha  \delta {\wek z\alpha}
- \wek {\dot z}\alpha \delta \kowek {\pi}\alpha -
\int _{\partial D} 
\wek z\alpha \delta \ten {\Theta}{\perp}\alpha  \ . 
\label {generFcal}
\ee
Now we have to complete the definition of the phase space
imposing the boundary conditions on $\ten {\Theta}{\perp}\alpha
\vert_{\partial D}$. The hamiltonian $F_D$ defines the
dynamics within the space {\bf R} defined this way. Such a dynamics
corresponds to the mixed Neumann--Cauchy problem for the field
equations. There are obviously other ways to control the boundary
conditions for our field theory. Each of them corresponds to
 a different hamiltonian but only the one corresponding to the
Dirichlet control mode is equal to the ``true'' energy of the
system. 

In the present paper we limit ourselves only to the discussion
of the Dirichlet control mode. In this case the generating
formula  (\ref{generHcal0}) is equivalent to the dynamical equations
(\ref{Hamilton2}) and (\ref{Hamilton1}).
Using equations (\ref{sofpi}), (\ref{vkofpi}) and the explicit form
of the hamiltonian density $H$ we can easily calculate the 
right
hand sides of these equations. We obtain
$$
\wek {\dot z}0 = -\frac 1\rho \ten z0k v^k + \frac T{\beta \rho} 
$$
$$
\wek {\dot z}a = -\frac 1\rho \ten zak v^k
$$
and
$$
\kowek {\dot \pi}0 = -\base xk (\beta s v^k )
$$
$$
\kowek {\dot \pi}a = \base xk [\rho \ten xka 
(f + pV - \frac 12 M\wek v2 - v^k \pi _a)]
$$
which is indeed equivalent to the field equations (\ref{r.pola4}).

At the end of this section let us observe that in the rest frame
(i.e. when $v^k=0$) the hamiltonian generating formula
reduces to the fundamental equation (\ref{de=}).
Indeed, in the rest frame we have $H=\rho e$, ${\dot z}^a =0$ and
therefore 
equation (\ref{generHinf}) takes the  following form:
$$
d(\rho e)=-(\kowek {\dot \pi}\alpha  + \base xk \ten
{\Theta}k\alpha ) \delta {\wek z\alpha} -
\ten {\Theta}k\alpha \delta \ten z{\alpha }k  
+ \wek {\dot z}\alpha \delta \kowek {\pi}\alpha =
$$
$$=
\wek {\dot z}0 \delta \kowek {\pi}0 -
\ten {\Theta}ka \delta \ten zak =
\frac T{\beta} \delta (\beta \rho s) +
(f+pV)\rho \ten xka \delta \ten yak =
$$
$$
= T\delta (\rho s) + (e - Ts+\frac p\rho )\delta \rho 
$$
which is obviously equivalent to the fundamental equation
(\ref{de=}) with both sides multiplied by $\rho$.

\section{Poisson bracket structure of thermo--hy\-dro\-dy\-namics}

The hamiltonian dynamics of the field theory can also be
equivalently formulated in terms of the Poisson bracket between
the physical observables i.e. functionals over the phase space 
{\bf P}. If $F$ and $G$ are two such functionals, their Poisson
bracket is defined by the standard formula (see \cite{Noether}):
\be
\{ F,G \} = \int _D
\frac{\delta F}{\delta \wek z{\alpha }(x)}
\frac{\delta G}{\delta \kowek {\pi }{\alpha }(x)} -
\frac{\delta G}{\delta \wek z{\alpha }(x)}
\frac{\delta F}{\delta \kowek {\pi }{\alpha }(x)} \ ,
\label{Poisson}
\ee
since the $\wek z{\alpha }$ and $\kowek {\pi }{\alpha }$ are
canonical variables i.e. their Poisson bracket is equal to the Dirac
delta distribution:
$$
\{ \wek z{\alpha }(x),\kowek {\pi }{\alpha }(y) \}= \delta (x-y) \
.
$$
The Poisson bracket of any physical observable $F$ with the
hamiltonian $H_D$ gives the time derivative of the
observable: \be
\dot F = \{ F,H_D\} \label{Heisenb.}
\ee
and therefore also the dynamical equations can be rewritten
this way:
$$
\wek {\dot z}\alpha =
 \{ \wek z\alpha ,H_D \} \ ,
$$
$$
\kowek {\dot \pi}\alpha =
\{ \kowek {\pi}\alpha , H_D \} \ .
$$
The reader may easily check that the above equations are
equivalent to equations (\ref{Hamilton1}) and (\ref{Hamilton2}).

The phase space {\bf P} can be factorized with respect to the
following equivalence relation: we call two elements of {\bf P}
equivalent if they have the same value of the physical parameters
$(v^k , \rho, s)$ or, equivalently,  $(p_k , \rho, s)$ (the
latter parameterization will be more suitable for some
calculations). Denote by {\bf Q} the corresponding quotient space
i.e. the space of the equivalence classes. It is interesting to
notice that the Poisson bracket structure can also be factorized to
{\bf Q}. Indeed, the formula  (\ref{Poisson}) implies the following
``generalized Poisson brackets'' for the quantities defined by
(\ref{sofpi}), (\ref{vkofpi}) and (\ref{ro}): 

$$
\{ p_k(x), p_l(y) \} =  \frac 2\rho 
\left[ p_k\base xl \delta (x-y)
- p_l\base xk \delta (x-y) \right]
$$
$$ 
\{ p_k(x), \rho (y) \} =  \base xk \delta (x-y)
$$
$$
\{ p_k(x), s(y) \} =  \frac s\rho \base xk \delta (x-y)
$$
$$
\{ s(x), \rho (y) \} = 0 \ .
$$
This means that the Poisson bracket of any two functionals which
are constant on equivalence classes will also be constant on
equivalence classes. Finally, we get the following formula for the
Poisson bracket of the two functions on {\bf Q}:
$$
\{F , G \} =  \int _D \int _D
\left(\frac{\delta F}{\delta p_k(x)}
\frac{\delta G}{\delta s(y)} -
\frac{\delta G}{\delta p_k(x)}
\frac{\delta F}{\delta s(y)}\right) \{ p_k(x), s(y) \} +$$
$$+
\left(\frac{\delta F}{\delta p_k(x)}
\frac{\delta G}{\delta \rho (y)} -
\frac{\delta G}{\delta p_k(x)}
\frac{\delta F}{\delta \rho (y)}\right)\{ p_k(x), \rho (y) \} +$$
$$+
\left(\frac{\delta F}{\delta p_k(x)}
\frac{\delta G}{\delta p_l(y)} -
\frac{\delta G}{\delta p_k(x)}
\frac{\delta F}{\delta p_l(y)}\right)\{ p_k(x), p_l(y) \} =$$
$$ = \int _D
  \frac {s(x)}{\rho (x)} \frac{\delta
F}{\delta p_k(x)} \base xk \frac{\delta G}{\delta s(x)} -
 \frac {s(x)}{\rho (x)}\frac{\delta G}{\delta p_k(x)}
\base xk \frac{\delta F}{\delta s(x)}  +
$$
$$
+  \frac{\delta F}{\delta p_k(x)}\base xk
\frac{\delta G}{\delta \rho (x)} -
 \frac{\delta G}{\delta p_k(x)}\base xk
\frac{\delta F}{\delta \rho (x)} +
$$
$$
+
 \frac {4p_k(x)}{\rho (x)} \frac{\delta F}{\delta
p_k(x)} \base xl\frac{\delta G}{\delta p_l(x)} -
\frac {4p_k(x)}{\rho (x)}\frac{\delta G}{\delta
p_k(x)} \base xl \frac{\delta F}{\delta p_l(x)}
 $$
(there are no surface integrals left after the integration by
parts because the elements of the configuration space satisfy the
boundary conditions).

Since the hamiltonian is also the functional on {\bf Q},
we can use the above ``generalized Poisson structure'' in order
to rewrite the
dynamical equations (\ref{Heisenb.}) in
terms of the physical observables, without any reference to the
original phase space {\bf P} (further reduction to the barotropic
case can be easily obtained if we limit ourselves to functions
which do not depend on the entropy).  Some authors (see e.g.
\cite{Zakharov}, \cite{Holm} or \cite{Levis-Marsden}) postulate
the above generalized structure {\it a priori}.
Our construction shows that in fact it is {\it the} canonical
structure of thermo--hydrodynamics and it does not depend upon
someone's lucky guesses.

We want to stress that ``reducing'' the phase space {\bf P} to
{\bf Q} is not a symplectic reduction typical for gauge theories.
The gauge transformations in the latter theories are defined as
``canonically conjugate to Hamiltonian constraints''. Therefore, the
future development of the system is determined by initial data up
to a gauge transformation only. Fixing a gauge at any instant of
time does not determine the trajectory unless we pass to the
quotient phase space.

Unlike in gauge theories, there are no
Hamiltonian constraints in {\bf P} and therefore
no symplectic reduction can
be performed. Formally, we can call gauge transformations the
unimodular reparameterizations of the material space and the
additive reparameterizations of the material time on each world
line of the fluid separately. Indeed, such a transformation
does not change the physical meaning of the initial data.
However, once fixed the initial gauge, the Cauchy problem for
canonical variables $(\wek z{\alpha} ,\kowek {\pi}\alpha )$ can 
be uniquely solved. This shows that
the canonical structure in {\bf P} is well adapted to the
initial value problem in the thermo--hydrodynamics and the
reduction to {\bf Q} is neither natural nor necessary.

\section{Complete Lagrange Picture}
 The dynamical history of the material can also be described in
terms of the mapping:  
$$
{\cal F} :Z\longmapsto M \ 
$$
inverse to the mapping ${\cal G}$ used in the Euler picture.

In this picture the laws of the thermo--hydrodynamics
will be expressed as
the second order partial differential equations for the field
variables $\wek x\mu (\wek y\alpha)$ which are the functions on the
"material" space--time $Z$. For this purpose we express
thermo--hydrodynamic quantities (the temperature, the velocity and
the density of the fluid) in terms of the first derivatives of the
fields. We have:
\be
T= \diff {\tau (t(\tau ,y^a),y^a)}t = (\ten x00 )^{-1} \ . 
\label{T L}
\ee
The definition of the inverse matrix
element:
$$
\ten x00 = \frac {\det (\ten zak )}{\det (\ten z{\alpha }{\mu })}
\ ,
 $$
together with equation (\ref{ro}), implies:
\be  
V = 
 (\ten x00 )^{-1} \det (\ten x{\mu}\alpha ) \ . \label{V L}
\ee 
Hence,
\be
\det (\ten x{\mu}\alpha ) = \frac VT = \frac 1{\rho T}
\ . \label{det4}
\ee
Moreover,
$$
0 =\diff {x^k}\tau (t(\tau ,y^a), y^a) = 
\diff {x^k}t (t, y^a) \diff t\tau (\tau ,y^a)  \ ,
$$
which implies:

\be
\wek vk=\frac {\ten xk0}{\ten x00}  \ . \label{vel L}
\ee
The above formulae express the velocity $v^k$, the molar volume $V$
and the temperature $T$ in terms of derivatives of the field
variables $x^\mu $.

Obviously, the numerical
value of the action integral corresponding to a given configuration
is the same in both the Lagrange picture and the Euler picture.
This means that the old lagrangian $L$ and the new lagrangian
${\cal L}$ are related via the following formula:
 $$ 
{\cal L} dz^0 \wedge dz^1 \wedge dz^2 \wedge dz^3 = 
{\cal F}^{\ast}(L dx^0 \wedge dx^1 \wedge dx^2 \wedge dx^3) =
$$
$$=  
 \det (\ten x{\mu}\alpha)  L dz^0 \wedge dz^1 \wedge dz^2
\wedge dz^3 = \frac {L}{\rho T}
dz^0 \wedge dz^1 \wedge dz^2
\wedge dz^3  
 $$
(unlike in the purely hydrodynamic case, when both the material and
the physical times coincide {\it a priori}, here the 4--dimensional
integrals have to be compared!). We conclude that the numerical
value of the new lagrangian equals: 
\be   
{\cal L}=\ten x00 (\frac 12
M\wek v2 - f(V,T)) \ ,\label{lagr L} 
\ee 
but now we have to express it in terms of the field variables 
$\wek x\mu $ and
their derivatives $\ten x{\mu}\alpha $ using equations (\ref{T L}),
(\ref{V L}) and (\ref{vel L}). Again, the Euler--Lagrange equations
together with the definition of the momenta $\ten {\cal
P}{\alpha}{\mu}$  canonically
conjugate to field  variables $x^\mu$ can be
interpreted as a generating formula for the symplectic relation:

\begin{equation} 
\delta {\cal L} = \base y{\alpha}(\ten {\cal P}{\alpha}\mu 
\delta \wek x\mu ) = 
(\base y{\alpha}\ten {\cal P}{\alpha}\mu) \delta \wek x\mu +
\ten {\cal P}{\alpha}\mu \delta \ten x\mu {\alpha} \ .
\label{sympl. rel.} 
\end{equation}
Using the explicit form of the lagrangian function and the
fundamental equation 
$$
df= -pdV -sdT
$$
 we obtain the following expressions
for the canonical momenta:
\be
\ten {\cal P}00 = \frac {pV}T \ten z00 - (e + pV + \frac 12 Mv^2 )
\label{pi0}
\ee

\be
\ten {\cal P}0k =  \frac {pV}T \ten z0k + M\kowek vk 
\label{pik}    
\ee

\be  
\ten {\cal P}a\mu = \frac {pV}T \ten za\mu 
\ ,  
\ee
with $\ten z{\alpha }\mu $ being a non--linear function of 
$\ten x{\mu }\alpha $, namely the element of its inverse matrix.

The dynamical equations of the theory imply the energy and momentum
conservation laws via the  N\"{o}ther theorem. The corresponding
energy--momentum tensor equals:

\begin{equation}
\ten {\cal T}\alpha \beta =\ten {\cal P}\alpha \mu \ten x\mu \beta
-\delta ^\alpha _\beta \ . {\cal L}  
\end{equation}

Using the explicit form of the lagrangian ${\cal L}$ we
obtain the following expressions for the components
of the energy--momentum tensor:

\begin{equation}
\ten {\cal T}00 = -s
\end{equation}
and therefore $-\ten {\cal T}00$ represents the entropy density,

\begin{equation}
\ten {\cal T}a0 = 0 \ ,
\end{equation}

\begin{equation}
\ten {\cal T}0a = Mv_k \ten xka - \ten x0a (e + pV +\frac 12 Mv^2)
\ , \end{equation}

\begin{equation}
\ten {\cal T}ab =\frac 1T \delta ^a_b (f + pV - \frac 12 Mv^2) 
\ . 
\end{equation}

Similarly as in the previous section, the zero--zero component of
the energy--momentum tensor plays the role of the hamiltonian of
the system with $\wek x{\mu}$ and $\Pi _{\mu }:=\ten {\cal P}0\mu
$ being canonical variables. To perform this Legendre
transformation we rewrite the generating formula (\ref{sympl. rel.})
as
follows:
\be
-\delta {\cal H}(\wek x\mu ,\ten x{\mu }a,\Pi _\mu ) 
= (\kowek {\dot \Pi}\mu  + \base za \ten {\cal P}a\mu )
\delta {\wek x\mu} +
\ten {\cal P}a\mu \delta \ten x{\mu }a  
- \wek {\dot x}\mu \delta \kowek {\Pi}\mu  \ ,
\label{generHinfL}
\ee
where the hamiltonian 
$$
{\cal H}:= \Pi _\mu \wek {\dot x}\mu - {\cal L} = 
\ten {\cal T}00 = -s 
$$
has to be expressed in terms of the canonical
variables. This can be done if we solve equations (\ref{pi0}) and
(\ref{pik}) with respect to
``velocities'' $\wek {\dot x}{\mu}$.
For this purpose we define:
$$
\chi := \det (\ten xka ) 
$$
and
$$
w_k := \ten {\tilde z}ak \ten x0a \ ,
$$
where by $\ten xka$ we denote the purely space--like
part of the matrix $\ten x{\mu}\alpha$ and by $\ten {\tilde z}ak$
we denote its 3--dimensional inverse. We stress that both the
scalar $\chi $ and the vector $w_k $ are defined
by the first
spacial derivatives of the configuration variables $\wek x{\mu}$.

Due to the
following  identities: 
$$
\chi = \frac VT \ten z00 \ ,
$$
$$
\frac{\ten z0k}{\ten z00} = -w_k
$$
we may rewrite definitions (\ref{pi0}) and
(\ref{pik}) in terms of the quantities $\chi$ and $w_k$:
\be
\kowek {\Pi}0 = p\chi - (e + pV + \frac 12 Mv^2 ) \label{1} \ ,
\ee

\be
\kowek {\Pi}k =   M\kowek vk - p\chi w_k  \label{2}   \ .
\ee
Moreover, we observe that the matrix 
$$
\ten xka - \frac {\ten xk0 \ten x0a}{\ten x00} =
\ten xka - \wek vk \ten x0a
$$
is equal to the 3--dimensional inverse of the matrix $(\ten yak )$.
Indeed:
$$
\left( \ten xka - \frac {\ten xk0 \ten x0a}{\ten x00} \right)
\ten yal = \delta ^k_l \ .
$$
Therefore
\be
V= \det (\ten xka - v^k \ten x0a ) = 
\chi \det (\delta ^k_l - v^k w_l ) =
\chi (1 - w_k v^k ) \ .
\label{3} 
\ee
From (\ref{1}), (\ref{2}) and (\ref{3}) we get
$$
MV= \chi (M - w_k\Pi ^k - p\chi w^2 )
$$ 
and
$$
-e= \Pi _0 + \frac {\Pi ^2}{2M} - \frac {p^2 w^2 \chi ^2}{2M} \ .
$$

Solving the above two equations with respect to $e$
and $V$ we can express them in terms of ``legal'' hamiltonian
variables, i.e. $\wek x{\mu}$ (together with its first
spacial derivatives)
and $\Pi _{\mu }$.

The above equations contain only two unknowns since 
for given constitutive equations (i.e. for a given fluid) the
pressure $p$ is a function of $e$ and $V$. The relation between
those quantities is given by the following fundamental equation:  
\be
ds(e,V) = \frac 1T de + \frac pT dV \ ,\label{ds=}
\ee
or, equivalently
$$
\frac 1T = {\frac {\partial s}{\partial e}}
$$
and
$$
p={\frac {\partial s}{\partial V}}
(\frac {\partial s}{\partial e})^{-1} \ .
$$
Finally, the hamiltonian of the system is equal to $-s(e,V)$ with
$e$ and $V$ expressed in terms of the canonical variables. 

Let us observe that in the rest frame (i.e. when $v^k=0$) the
hamiltonian generating formula  reduces to the
fundamental equation (\ref{ds=}). Indeed, in the rest frame we
have $\chi = V$, $\Pi _0 = e$ and $\wek {\dot x}k = 0$.
Therefore, the equation (\ref{generHinfL}) reduces to
$$
\delta s 
= 
\frac {pV}{T}\ten za\mu \delta \ten x{\mu }a  
- \wek {\dot x}0 \delta \kowek {\Pi}0 =
p \left( \frac {V}{T}\ten z{\alpha}\mu \delta \ten x{\mu }{\alpha} -
\frac {V}{T}\ten z0\mu \delta \ten x{\mu }0 \right) +
 \frac {1}{T}\delta e =
$$
$$
=
p\left( \delta (\frac {V}{T}) - V\delta (\frac {1}{T}) \right) +
\frac {1}{T}\delta e =
\frac {p}{T}\delta V + \frac {1}{T}\delta e 
\ .
$$

The
equation (\ref{ds=}) has been obtained from (\ref{de=}) dividing
both sides by $T$. Both (\ref{de=}) and (\ref{ds=}) can be 
interpreted as symplectic relations between
thermo--hydrodynamic quantities characterizing the material. In
the first case the energy $e=e(s,V)$ is a generating function of a
lagrangian submanifold in the symplectic space of parameters
$(s,V,T,p)$, equipped with the symplectic
form  $\ dT \wedge ds + dV \wedge dp \ $ and with $(s,V)$ chosen as
control parameters. In the second case the same physical relation is
described by the generating function $s=s(e,V)$ in the symplectic
space of parameters $(e,V,\frac 1T , \frac pT )$, equipped with the
symplectic form $\ d(\frac 1T ) \wedge de + d(\frac pT ) \wedge
dV\ $ and with $(e,V)$ chosen as control parameters. The
transformation between those two formulations of thermodynamics,
called respectively the ``energy picture'' and the ``entropy
picture'', is rather obscure. It has nothing to do with the Legendre
transformations (e.g. the Helmholtz transformation leading from
(\ref{de=}) to (\ref{sdT}) and similar transformations relating the
energy to the enthalpy or to the Gibbs free energy). Legendre
transformations consist always in exchanging a control parameter
(e.g. the entropy) with the corresponding ``response parameter'' (e.g.
the temperature).  They always refer to a fixed lagrange submanifold within a
fixed phase space, which is described with respect to two different
``control modes'' (the comprehensive description of these phenomena
can be found in  \cite{Kij-Tulcz}).
 
Unfortunately, the transformation mixing the phase space parameters
with the generating function itself is of different nature and we
do not know in the literature any convincing interpretation of
such a transformation. Here, we obtain such an interpretation in a
natural way: 
there are two possible descriptions of the dynamics depending on
whether we parameterize the physical events by the physical time or
by the material time. The transformation between these two
parameterizations exchange also the role of the momenta canonically
conjugate with both times (the energy and the entropy).
  
\section{Relativistic theory}

The relativistic version of the above theory can be found in
\cite{KSG}. Here we give the short review of the results.

 In Euler
picture the relativistic thermodynamics is a field theory with four
field potentials $z^{\alpha }=z^{\alpha }(x^{\mu })$ defined on the
physical space--time $M$ equipped with the metric structure $g_{\mu
\nu }$ (not necessarily flat!). The structure of the material
space--time remains the same as in non--relativistic case.
Similarly, we introduce the matter current 
$$
j:={\cal G}^*(r)
$$
which is automatically conserved since
$$
dj=d{\cal G}^*(r)={\cal G}^*(dr)=0 \ .
$$
In terms of coordinates the components of $j$ are given by 
the formula
$$
j^{\kappa }:=\epsilon ^{\kappa \mu \nu \lambda } 
\ten z1{\mu } \ten z2{\nu } \ten z3{\lambda }
$$
which implies the continuity equation
$$
\base x\mu j^{\mu } =0  \ .
$$
Decomposing now
$$
j^{\mu } =\sqrt{\det (g_{\mu \nu })}\ \rho u^{\mu } \ ,
$$
where the velocity vector
 $u^{\mu }$ is normalized, we obtain the definition of the
hydrodynamic quantities $(\rho ,u^{\mu })$ in terms of the first
derivatives of the three ``spatial'' field potentials $z^k$ (we
stress that $\rho $ is now the {\it rest frame density}). The
fourth potential is used again to define the temperature as a
ratio between the two times running along the world lines of the
fluid:
$$
T:= \beta u^{\mu } \ten z0\mu \ .
$$
We choose the rest frame free energy density as a Lagrangian of
the theory:
$$
L=L(\ten z{\alpha }{\mu })= - \rho f_{rel.}(\rho , T) \ .
$$
There is no kinetic energy term since
the relativistic free energy
$f_{rel.}$ contains also the rest mass contribution to the energy:
$$
f_{rel.}:=f+M
$$
(we use geometric units with the speed of light being equal to 1).
The Euler-Lagrange equations together with the definition of
the momenta $\ten {\Theta}{\mu}{\alpha}$ canonically conjugate to
field  variables $z^\alpha$ can be
written as a generating formula for the symplectic relation:

$$ 
\delta L(\wek z\alpha ,\ten z{\alpha}{\mu} ; x^\mu )
= \base x{\mu}(\ten {\Theta}{\mu}\alpha \delta \wek z\alpha ) = 
(\base x{\mu}\ten {\Theta}{\mu}\alpha) \delta \wek z\alpha +
\ten {\Theta}{\mu}\alpha \delta \ten z{\alpha}{\mu} \ .
$$
To check that the above field theory really describes the
thermo--hydrodynamics we calculate the components of the energy
momentum tensor given again by the formula (\ref{tmunu}). After
some calculations we obtain:
$$
\ten t{\mu }{\nu } = - \sqrt{\det (g_{\mu \nu })}\ \left(
\rho e u^\mu u_\nu + p({\delta }^{\mu }_{\nu } + u^\mu u_\nu )
\right) \ .
$$
The N\"other theorem now reads:
$$
\nabla _\mu \ten t{\mu }{\nu } = 0
$$
which, together with the entropy equation, is really equivalent to
the equations of thermo--hydrodynamics.

The hamiltonian formulation and canonical structure of the theory
can be obtained from the construction very much similar to the
non--relativistic one.

To describe the interaction of the hydrodynamic matter with the
gravitational field it is sufficient to add the gravitational
lagrangian to the above matter lagrangian and to take into account
the gravitational degrees of freedom, describing the geometry of
$M$. Such a theory of a self--gravitating fluid becomes especially
simple in the complete Lagrange picture.The  main advantage of this
picture consists in the fact that the ``gravitational gauge'' and
the ``hydro--thermodynamic gauge'' eliminate somehow each other
and the theory can be formulated in terms of purely physical (gauge
free!) quantities. This way we obtain the theory which is well
adapted to the numerical analysis and can be used e.g. to computer
simulations of the geometrodynamics in both cosmology and
astrophysics  (see again \cite{KSG}).

Another advantage of our approach consists in a rather
straightforward generalization of the hydrodynamics to the
elastodynamics. The relativistic formulation of the latter has
never obtained any satisfactory formulation. Here, the
introduction of the riemannian structure into the material space
$Y$ enables us to generalize the theory in such a way that it is
relativistic and reduces in the rest frame to the
non--relativistic, nonlinear elastodynamics. The paper containing
these results will be soon published.

\section{Microscopic\  \ interpretation\ \ of\ \ the\ \ material\ \
time}

We suppose that the liquid is composed of molecules with mass $m$. 
If the temperature of the fluid equals $T$, the molecules move
chaotically around the theoretical world lines of the fluid (lines
tangent to the vector field  $\wek u\mu$) and the mean kinetic
energy of this motion with respect to the rest frame is equal (for
 the low temperature) to $\frac 32 kT=\frac 12 m\wek v2$. Due to this
motion the proper time $t$ for the particles is retarded with
respect to the physical time $x^0$ calculated along $\wek u\mu$.
For velocities $v$ much smaller than the velocity of light $c$ this
retardation can be calculated from the formula:
$$
	 t =\int \sqrt{1 - \frac{\wek v2}{\wek c2}}\ dx^0 \approx  
(1-\frac{\wek v2}{2\wek c2}) x^0 = x^0 - x^0\frac{3kT}{2mc^2} \ . 
$$
We identify the parameter $\tau$ with the proper time retardation:
$$
	\tau:= x^0 - t = \frac{1}{\beta}  x^0 T	 \ ,
$$
where the constant $\beta =\frac{2mc^2}{3k}$ has the dimension of
temperature. Hence,
$$
	\beta \diff{\tau}{x^0}  = T	
$$
similarly as in formula (\ref{deftemp}). We interpret therefore
the "material time" as
the "proper time retardation" due to the chaotic motion of the
particles. This phenomenon enables us to construct (at least
theoretically) a "radium thermometer". We inject a drop of the
radioactive radium into the fluid. Due to the chaotic motion of the
particles the lifetime of the radium gets lengthened proportionally
to the temperature of the fluid. Therefore, measuring the lifetime
we measure the temperature.


\begin{thebibliography}{666}


\bibitem{Noether}N. N. Bogoliubov, D. Shirkov, {\it Introduction into
Theory of Quantum Fields} (Science, Moscou, 1976);
 E. N\"other (1918)


\bibitem{Chernoff-Marsden} P. R. Chernoff and J. E. Marsden
{\it Properties of infinite dimensional hamiltonian systems},
Springer Lecture Notes in Mathematics, vol. 425, 1974

  
\bibitem{KSG} 
A. G\'ornicka, J. Kijowski and A. Sm\'olski, {\it Hamiltonian theory
of self--gravitating perfect fluid and a method of effective
deparameterization of Einstein's theory of gravitation}, to appear
in Phys. Rev. D 


\bibitem{Holm} D. D. Holm; Physica {\bf 17D} (1985) p. 1--36


\bibitem{ComptesRendues} J. Jezierski and J. Kijowski, Comptes
Rendues Acad. Sci. Paris, serie II, 301 (1985) p. 221-224


\bibitem{KPT} J. Kijowski, B. Pawlik, W. M. Tulczyjew, Bull.
Acad. Polon. Sci. (math. phys. astr.) 27 (1979) p. 163; H.P.
K\"unzle, J. M. Nester, J. Math. Phys. 25 (1984) p. 1009

\bibitem{Kij-Tulcz2} J. Kijowski and W. M. Tulczyjew, {\it
Relativistic hydrodynamics of isoentropic flows}, Mem. Acad. Sci.
Torino,serie 5, No.6-7 (1982-83) p. 3-17 

\bibitem{Kij-Tulcz} J. Kijowski and W. M. Tulczyjew, {\it A
Symplectic Framework for Field Theories}, Lecture Notes in Physics
vol.107, Springer-Verlag, Berlin, (1979)


\bibitem{Levis-Marsden} D. Levis, J. Marsden, R. Montgomery;
Physica, {\bf 18D} (1986) p. 391--404 
 
\bibitem{Tulcz} W. M. Tulczyjew, {\it Hamiltonian systems, Lagrangian
systems and the Legendre transformation}, Symposia matematica 14
(1974) p. 247; M. R. Menzio and W. M. Tulczyjew, {\it
Infinitesimal symplectic relations and generalized Hamiltonian
dynamics}, Ann. Inst. Henri Poincare, 28 (1978) p. 349

\bibitem{Zakharov}V. E. Zakharov and E. A. Kuznetsov {\it
Hamiltonian formalism for systems of hydrodynamic type}, Soviet
Scientific Review, Section C: Mathematical Physics Review, {\bf
4} (1984) p.167--219 

 
\end{thebibliography}
 \end{document}